\documentclass[journal]{IEEEtran}

\usepackage{adjustbox,lipsum}
\usepackage{cite}
\usepackage{float}
\usepackage{multicol}
\usepackage{multirow}
\usepackage{booktabs}
\usepackage{makecell}
\usepackage{algorithm}
\usepackage{amssymb}
\usepackage{algpseudocode}
\usepackage{graphicx}
\usepackage{textcomp}
\usepackage{xcolor}
\usepackage{textcomp}
\usepackage{amsmath}
\usepackage{array}
\usepackage{url}
\def\BibTeX{{\rm B\kern-.05em{\sc i\kern-.025em b}\kern-.08em
    T\kern-.1667em\lower.7ex\hbox{E}\kern-.125emX}}

\usepackage{color,soul}

\usepackage{threeparttable}

\begin{document}


\title{Optimal Utilization of Third-Party Demand Response Resources in Vertically Integrated Utilities: A Game Theoretic Approach}

\author{Sayyid Mohssen Sajjadi,~\IEEEmembership{Student Member,~IEEE,}
        and Meng Wu,~\IEEEmembership{Member, IEEE}
\thanks{The authors are with the School of Electrical, Computer and Energy Engineering, Arizona State University, Tempe, AZ, 85281 USA e-mail: (ssajjad1@asu.edu, mwu@asu.edu).}}


\maketitle

\begin{abstract}
This paper studies the optimal mechanisms for the vertically integrated utility to dispatch and incentivize the third-party
demand response (DR) providers in its territory. A framework is proposed, with three-layer coupled Stackelberg and simultaneous games, to study the interactions and competitions among the profit-seeking process of the utility, the third-party DR providers, and the individual end users (EUs) in the DR programs. Two coupled single-leader-multiple-followers Stackelberg games with a three-layer structure are proposed to capture the interactions among the utility (modeled in the upper layer), the third-party DR providers (modeled in the middle layer), and the EUs in each DR program (modeled in the lower layer). The competitions among the EUs in each DR program is captured through a non-cooperative simultaneous game. An inconvenience cost function is proposed to model the DR provision willingness and capacity of different EUs. The Stackelberg game between the middle-layer DR provider and the lower-layer EUs is solved by converting the original bi-level programming to a single-level programming. This converted single-level programming is embedded in an iterative algorithm toward solving the entire coupled games framework. Case studies are performed on IEEE 34-bus and IEEE 69-bus test systems to illustrate the application of the proposed framework.

\end{abstract}


\IEEEpeerreviewmaketitle

\section{Introduction}

\IEEEPARstart{P}{rice}-responsive prosumers, including distributed energy resources and flexible consumers (such as air conditioners with smart thermostats), can be aggregated by third-party demand response (DR) providers which offer grid services to both deregulated wholesale markets (operated by independent system operators, ISOs) and vertically integrated utility companies (UCs). Different from ISOs which follow systematic market clearing mechanisms to dispatch and incentivize DR providers, utilities' procedure for dispatching/incentivizing these third-party DR providers tend to be heuristic and unclear. To design appropriate mechanisms for UCs to harness DR services, 1) interactions among the UC, the (strategic) third-party DR providers in the UC's territory, and the (strategic) end-user (EU) prosumers in each DR provider's territory need to be studied; and 2) each EU's willingness for DR provision needs to be considered. The ideal mechanism should consider DR benefits for all parties by providing optimal dispatch/price signals from the UC to the third-party DR providers, and from the DR providers to the EUs. 

There are existing works addressing the above issues. References \cite{ellman2020incentives,mohandes2020incentive} design DR programs that incentivize EUs to participate in the DR program and maximize their profit. The work in \cite{ellman2020incentives} only models EUs' profit maximization objectives in the DR program, without considering decision making process of the DR provider or the UC when dispatching/incentivizing individual DR resources of EUs. The DR program in \cite{mohandes2020incentive} does not model inconvenience cost of the EUs which determines the EU's willingness for DR provision. DR programs based on baseline mechanism are proposed in \cite{muthirayan2019mechanism,muthirayan2019minimal} that penalize the EUs whose power consumption deviates from the reported baseline. These baseline mechanisms in \cite{muthirayan2019mechanism,muthirayan2019minimal} select EUs randomly during DR events. In an efficient DR program with many EUs, it is essential that the DR mechanism selects EUs during a DR event based on the EUs' capability and willingness of DR provision at that moment \cite{PES}. The DR program in \cite{tsaousoglou2020truthful,bahrami2020deep,jia2016dynamic,liu2017energy} takes the EUs' discomfort levels into consideration. However, the willingness of different EUs is assumed to be identical, which is unrealistic. Also, it is shown in \cite{liu2017energy} that their solution methods diverge when the willingness coefficient is small. A distributed DR control mechanism based on the Lyapunov optimization is proposed in \cite{zheng2014distributed} to dispatch controllable loads in residential EUs aiming to alleviate the fluctuations of the intermittent renewable energy sources. In \cite{vivekananthan2014stochastic}, a stochastic ranking algorithm is proposed to control the thermostatically controllable household appliances and provide regulation services to the grid.
A heuristic DR program is proposed in \cite{parizy2018low} that utilizes a hopping scheme to schedule the controllable loads. The advantage of the proposed DR program in \cite{parizy2018low} is that it improves the EUs' privacy because it does not require two-way communication between the EUs and the DR operator. However, the EUs' preferences of maximizing their own benefits/happiness during a DR event are not considered in \cite{parizy2018low,vivekananthan2014stochastic,zheng2014distributed}, which may discourage EUs for DR provision.

As DR programs are widely adopted in regulated UCs, it is vital to understand the behavior and interactions of different entities in the DR program (such as the EUs in the residential/business DR programs, the third-party DR providers offering aggregated DR services, and the UC utilizing these aggregated DR services), since their behavior/interactions could greatly impact the overall efficiency of the DR program \cite{ellman2020incentives}. Game theory is a prominent tool in modeling the interactions among these DR entities \cite{yu2015real,jia2016dynamic,maharjan2013dependable,nekouei2014game,tavakkoli2020bonus,zhang2020stochastic,lu2018data,hurtado2017enabling,ma2015cooperative,collins2017distributed,barabadi2019new}. In \cite{yu2015real,jia2016dynamic,maharjan2013dependable,nekouei2014game,tavakkoli2020bonus,zhang2020stochastic,lu2018data}, the bi-level Stackelberg game is adopted to study such interactions. These works ignore the interactions between the UC and multiple DR providers for exchanging DR services and compensations. A non-cooperative game is adopted in \cite{hurtado2017enabling,collins2017distributed, ma2015cooperative, barabadi2019new} that models the competition among the EUs. These works do not offer compensation to the EUs in return for their contribution in the DR program. This may discourage the EUs for DR provision. The DR program in \cite{collins2017distributed, barabadi2019new} also do not consider discomfort levels of the EUs in the DR program. A three-stage game model is proposed in \cite{zhou2017game} where the UC and the energy storage company in the upper layer set price signals to the microgrid. The microgrid in the middle layer then adjusts its energy price to the EUs. Finally, the EUs in the lower layer update their power consumption as a response to the prices from the microgrid. However, this work does not model the interactions among EUs within the microgrid. In a competitive DR program, EUs could compete with each other to maximize their profit. Each EU's strategy in response to other EUs' actions can be modeled via a simultaneous game.

Several challenges remain unexplored in existing literature: How could the UC optimally incentivize the profit-seeking individual EUs and third-party DR providers for their DR service provision/aggregation? How to incorporate competition among individual EUs into the framework for studying interactions among the UC, the DR providers, and the EUs? How to incorporate different willingness of individual EUs for DR provision in the framework for studying such interactions?

This paper is built upon our prior work in \cite{PES} which presents a simplified model to study the interactions among different entities in the third-party DR programs for the regulated UC. Major contributions of this paper are as follows:
\begin{itemize}
\item A framework is proposed for the regulated UC to optimally dispatch and incentivize aggregated DR resources provided by third-party business and residential DR providers, considering profit maximization of the UC, the third-party DR providers, and individual EUs. Interactions among different entities are studied through coupled Stackelberg and simultaneous games.
\item The proposed framework models the interactions between the UC and the third-party business/residential DR providers by a Stackelberg game which is solved using an iterative method.
\item The proposed framework models the interactions between the third-party business/residential DR providers and the business/residential EUs by another Stackelberg game. The model is formulated as an optimization problem constrained by other optimization problems (OPcOP) and is solved by converting the bi-level problem to a single-level problem.
\item The competition among the EUs in the same DR program is modeled by a non-cooperative simultaneous game.
\item An inconvenience cost function is proposed to model the willingness and capability of each EU for DR provision.
\end{itemize}

To our best knowledge, this is the first attempt studying optimal DR dispatch and incentivization for the regulated UC, considering comprehensive interactions among the UC, the strategic third-party business/residential DR providers, and the competitive EUs with different DR provision willingness. The rest of the paper is organized as follows. Section II presents the structure of the framework with coupled games. Section III formulates the optimization problems in the coupled games. Section IV presents the solution approach. Section V presents the case study results. Section VI concludes this paper.

\section{Proposed Framework with Coupled Games}
\begin{figure}[!t]
	\centering
	\includegraphics[width=0.42\textwidth]{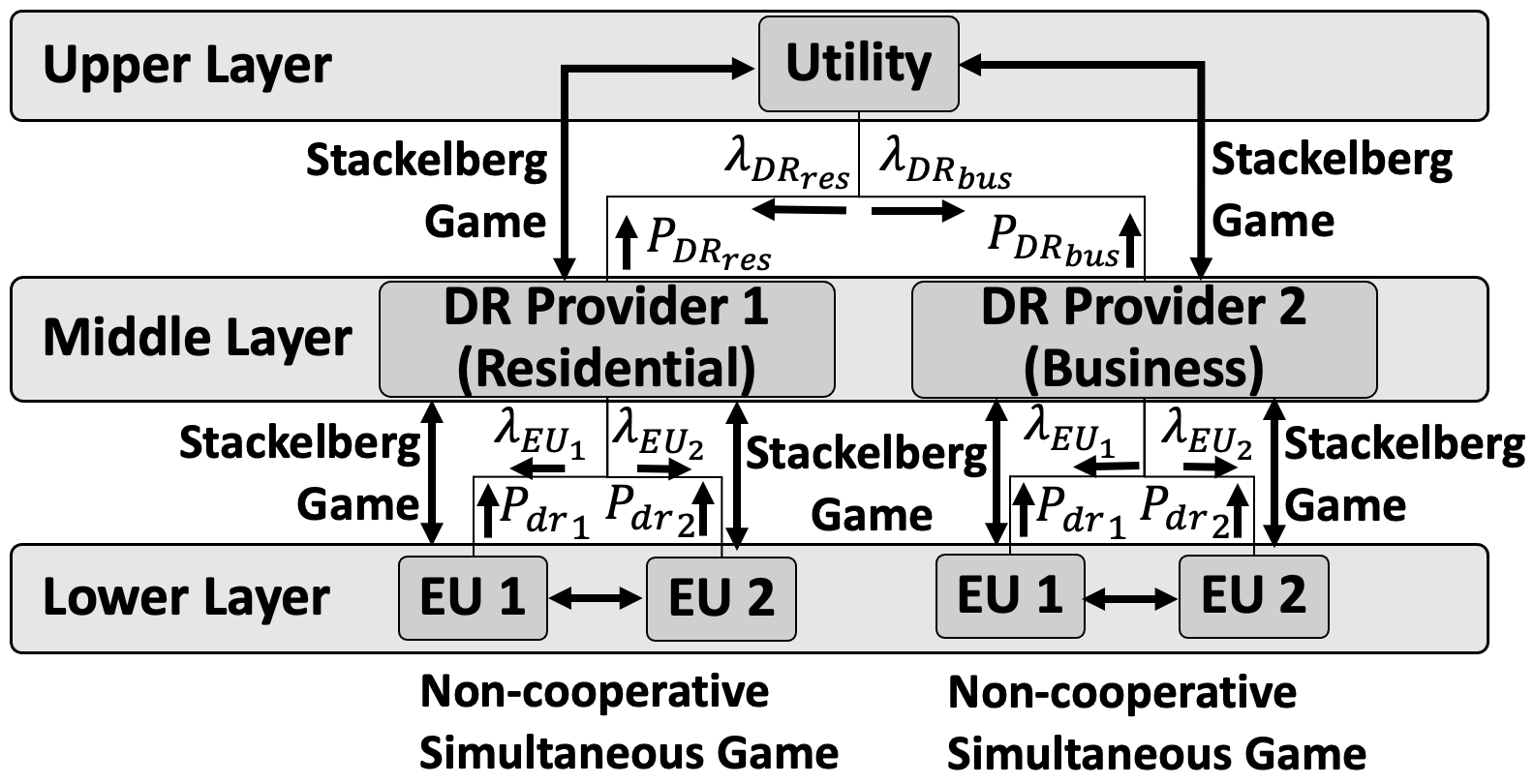}
	\caption{Structure of the proposed framework with coupled games.}
	\label{Schematic Diagram}
\end{figure}

Fig.~\ref{Schematic Diagram} shows the proposed framework, with the coupled Stackelberg games and non-cooperative simultaneous games to capture the interactions among the UC, the business/residential DR providers, and the business/residential EUs. The UC is the major leader of the three-layer game. The Stackelberg game between the UC (leader) in the upper layer and the DR providers (followers) in the middle layer is coupled with another Stackelberg game between the DR providers (leaders) in the middle layer and the EUs (followers) in the lower layer. These coupled Stackelberg games allow the UC, the DR providers, and the EUs to jointly determine the optimal prices the UC pays the DR providers for DR management, the optimal prices the DR providers pay the EUs for DR provision, and the optimal DR provision of each EU in the DR programs, respectively, considering profit maximization of all the above entities. The simultaneous game in the lower layer captures the competition among the EUs in the same DR program. Each EU reports a DR provision willingness parameter (a scalar between 0 and 1) to the corresponding DR provider. The DR providers run the simultaneous game among the EUs, considering the EUs' DR provision willingness when determining their optimal DR quantity provision. The simultaneous game among the EUs is non-cooperative. If an EU's DR provision willingness increases, the EU will contribute more in the DR program and earn more profit, which may reduce the DR provision quantity and profit of other EUs.

\section{Problem Formulation}
This section formulates the optimization problems in the coupled games in Fig.~\ref{Schematic Diagram}.

\subsection{Lower Layer Optimization Problem}
The lower layer problem models the decision making process of competitive EUs in each DR program via the simultaneous game. Each EU maximizes its profit considering inconvenience cost and the revenue for DR quantity provision. The simultaneous game captures the EUs' behavior against other EUs' best response. The proposed simultaneous game does not randomly select the EUs to contribute in the DR program \cite{muthirayan2019mechanism,muthirayan2019minimal}. Instead, EUs are selected considering each EU's revenue/inconvenience for DR provision.

\subsubsection{Inconvenience Cost}
The inconvenient cost reflects the EU's willingness to curtail or shift load during a DR event. Each EU's inconvenience cost as a function of its DR quantity provision is modeled as follows.
\begin{align}
	C_{{Inc}_{ij}}^t(P_{{dr}_{ij}}^t)=\frac{P_{{dr}_{ij}}^t}{P_{{dr-max}_{ij}}^t-P_{{dr}_{ij}}^t}   & & \forall t \in T, \forall i \in I, \forall j \in J
	\label{eqn:fitting function re-written - 3}
\end{align}
where
\begin{align}
	P_{{dr-max}_{ij}}^t=\alpha_{ij} \cdot P_{b_{ij}}^t, \qquad \alpha_{ij} \in [0,1] 
	\label{eqn:alpha}
\end{align}
where $t$ and $T$ denote the index and set for time intervals of a DR event, respectively; $i$ and $I$ denote the index and set for the third-party DR providers in the UC territory, respectively; $j$ and $J$ denote the index and set for the EUs in the same DR program, respectively; $C_{{Inc}_{ij}}^t(\cdot)$, $P_{{dr}_{ij}}^t$, and $P_{{dr-max}_{ij}}^t$ denote the inconvenience cost, the DR quantity provision, and the upper bound for DR quantity provision of $j^{th}$ EU within $i^{th}$ DR program at time $t$, respectively; $\alpha_{ij} \in [0,1]$ denotes the $j^{th}$ EU's willingness parameter reported to the $i^{th}$ DR provider during a DR event; $P^t_{b_{ij}}$ denotes the base power consumption of $j^{th}$ EU within $i^{th}$ DR program at time $t$.
\begin{figure}[!t]
	\centering
	\includegraphics[width=0.27\textwidth]{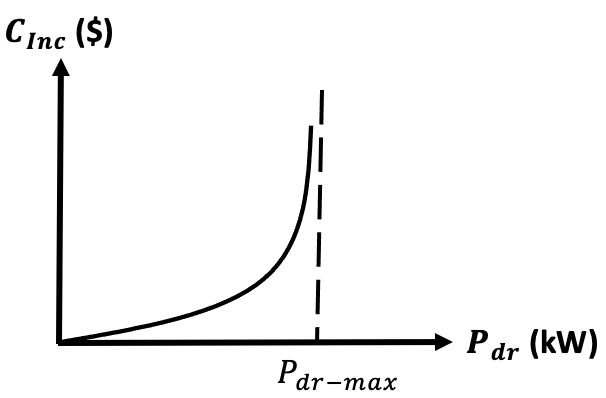}
	\caption{Inconvenience cost of an EU as a function of its DR quantity provision.}
	\label{Inconvenience Cost}
\end{figure}
The inconvenience cost function in (\ref{eqn:fitting function re-written - 3}) is shown in Fig.~\ref{Inconvenience Cost}. The inconvenience/discomfort level of an EU increases from $0$ to $+\infty$ as the amount of DR quantity provision (achieved by curtailing/shifting loads) increases from $0$ to the upper bound. In (\ref{eqn:alpha}), this upper bound is determined by both the EU's base power consumption (i.e., the EU's maximum capability for DR provision) and its willingness for DR provision (i.e., $\alpha_{ij}$).

\subsubsection{Revenue for DR Quantity Provision}
The DR provider pays the EUs for purchasing their DR quantity. The revenue of $j^{th}$ EU within $i^{th}$ DR program at time $t$ is modeled by $R^t_{{EU}_{ij}}$ as follows.
\begin{align}
    R^t_{{EU}_{ij}} = \lambda_{EU_{ij}}^t \cdot P_{{dr}_{ij}}^t   & & \forall t \in T, \forall i \in I, \forall j \in J
\label{eqn:EU Revenue}
\end{align}
where $\lambda^t_{{EU}_{ij}}$ denotes the DR price $i^{th}$ DR provider pays $j^{th}$ EU for DR quantity provision.

\subsubsection{Optimal Decisions of the EUs}
The optimization problem for $j^{th}$ EU within $i^{th}$ DR program can be modeled as follows.
\begin{equation}
\begin{aligned}
    \smash{\displaystyle \max_{P_{{dr}_{ij}}^t \in [0, P_{{dr-max}_{ij}}^t]}} \; \quad \sum \limits_{t \in T} \left[\lambda_{EU_{ij}}^t \cdot P_{{dr}_{ij}}^t - C_{{Inc}_{ij}}^t(P_{{dr}_{ij}}^t) \right] \\
    \forall i \in I, \forall j \in J
\label{OF_EU}
\end{aligned}
\end{equation}

Equation (\ref{OF_EU}) maximizes each EU's profit, considering the raw DR revenue in (\ref{eqn:EU Revenue}) and inconvenience cost in (\ref{eqn:fitting function re-written - 3}). The DR quantity provision of each EU at time $t$ is constrained within its lower bound and upper bound. Each EU determines its optimal DR quantity provision in response to the price received from the DR provider, considering the simultaneous game/competition among the EUs. Further discussion on the simultaneous game/competition among the EUs is provided in next section where the optimization problem for middle layer entities is presented.

\subsection{Middle Layer Optimization Problem}
The DR providers in the middle layer maximize their profit by adjusting their prices to the EUs, in response to the price they receive from the UC and the DR quantities they receive from individual EUs. The $i^{th}$ DR provider's optimization problem is formulated as follows.
\begin{align}
    \smash{\displaystyle \max_{\lambda^t_{{EU}_{ij}} \in [0,\lambda^t_{DR_i}]}} \; \sum \limits_{t \in T} \sum \limits_{j \in J} (\lambda^t_{{DR}_i} - \lambda^t_{{EU}_{ij}}) \cdot P_{{dr}_{ij}}^t    & & \forall  i \in I
\label{OF - ML}
\end{align}
where $\lambda^t_{{DR}_i}$ denotes the DR price the UC pays $i^{th}$ DR provider for managing the DR program. Each DR provider maximizes its profit, considering its revenue from the UC and its cost for paying individual EUs. 

In the simultaneous game/competition among the EUs, if an EU is willing to contribute more in the DR program, the aggregated DR quantity the DR provider purchases increases. Providing more DR quantity from the DR provider to the UC encourages the UC to decrease the price incentive $\lambda_{{DR}_i}^t$. This results in decreasing the optimal price incentive $\lambda^t_{{EU}_{ij}}$ (see (\ref{OF - ML})), which impacts the rest of the EUs within the same DR program and may reduce their DR provision quantity and profit.

\subsection{Upper Layer Optimization Problem}
The UC sends DR price signals to the DR providers and receives aggregated DR quantity. The UC's objective is to maximize its profit considering operation cost for system-wide non-DR resources, the EUs' electricity bills, and the payment to the DR providers. The mathematical formulation of each component is as follows.

\subsubsection{Operation Cost Reduction}
The following quadratic cost function is adopted to model the total cost of operating the UC system using non-DR generating resources at time $t$.
\begin{align}
    C_{g}^t(P_{g}^t) = c_0 + c_1\cdot P_{g}^t + c_2 \cdot (P_{g}^t)^2  & &   \forall t \in T
    \label{eqn_10}
\end{align}
where $C_{g}^t(\cdot)$ and $P_{g}^t$ are the total operating cost and the total power output of all the non-DR generating resources across the UC's footprint at time $t$, respectively; $c_0$, $c_1$ and $c_2$ are the averaged constants of the system-wide quadratic generation cost function \cite{PES}.

During a DR event, DR resources are called to reduce total power supplied by non-DR resources. The total operating cost for non-DR resources for $\forall t \in T$ is reduced as follows \cite{tsaousoglou2020truthful}.
\begin{equation}
\begin{split}
	\Delta C_{g}^t &= C_{g}^t(P_{g-pre}^t) - C_{g}^t(P_{g-post}^t) \\
	&=  C_{g}^t(P_{g-pre}^t) - C_{g}^t(P_{g-pre}^t - \sum \limits_{i \in I} P_{{DR}_i}^t) \\
	&= (c_1+2\cdot c_2 \cdot P_{g-pre}^t) \sum \limits_{i \in I} P_{{DR}_i}^t - c_2 \left(\sum \limits_{i \in I} P_{{DR}_i}^t\right)^2 \\ 
\end{split}
	\label{OCR}
\end{equation}
where at time $t$, $P_{g-pre}^t$ and $P_{g-post}^t$ are the total power supplied by all non-DR generating resources without and with calling the DR event, respectively; $P_{{DR}_i}^t$ is the aggregated DR quantity from $i^{th}$ DR provider. We have $P_{g-post}^t=P_{g-pre}^t - \sum \limits_{i \in I} P_{{DR}_i}^t$ and $P_{{DR}_i}^t=\sum \limits_{j \in J} P_{{dr}_{ij}}^t$.

During a DR event, equation (\ref{OCR}) models the total operation cost reduction of system-wide non-DR generating resources as a function of the aggregated DR quantity.

\subsubsection{Revenue from EUs' Electricity Bills}
This component considers the UC's revenue for the electricity bills received from the EUs during the DR event. The total revenue earned from the electricity bills at time $t$ during a DR event is modeled by $R_{EB}^t$ as follows.
\begin{align}
R_{EB}^t & = \sum \limits_{i \in I} \lambda^t_{ret_i} \cdot (P^t_{B_i} - P_{{DR}_i}^t)   & & \forall  t \in T
\label{Revenue - EB}
\end{align}
where $P^t_{B_i}=\sum \limits_{j \in J} P_{b_{ij}}^t$ denotes the total base load at time $t$ (without the DR event) in $i^{th}$ DR program; $\lambda^t_{ret_i}$ denotes the retail rate at time $t$ for the EUs in $i^{th}$ DR program. The retail rates for residential EUs and business EUs are different.

\subsubsection{Payments to the DR Providers for Aggregated DR Provision}
This cost for the UC is modeled as follows.
\begin{align}
C_{UC}^t = \sum \limits_{i \in I} \lambda^t_{DR_i} \cdot P_{{DR}_i}^t & & \forall  t \in T
\label{payment -UC}
\end{align}
where $C_{UC}^t$ is the total payment at time $t$ from the UC to all DR providers for offering aggregated DR quantities.

\subsubsection{Optimal decision of the UC}
The UC's profit maximization problem is modeled in (\ref{OF Utility}), respectively.
\begin{equation}
\begin{aligned}
    \smash{\displaystyle \max_{\lambda^t_{{DR}_i} \geq 0}} \; R_{UC} = \sum \limits_{t \in T} & \left( R_{EB}^t - C_{UC}^t + \Delta C_g^t \right)
\end{aligned}
\label{OF Utility}
\end{equation}
where $R_{UC}$ is total profit the UC earns during the DR event. In ({\ref{OF Utility}}), the UC adjusts its price signals to the DR providers in response to the aggregated DR quantities it receives from the DR providers such that the UC's profit considering the EUs' electricity bills, the UC's payments to the DR providers, and the operation cost of the non-DR resources is maximized.

\section{Solution method}
The Stackelberg game between each middle-layer DR provider (leader) and the lower-layer EUs (followers) is formulated as an OPcOP and is solved by converting the bi-level problem into a single-level problem \cite{gabriel2012complementarity}. The Stackelberg game between the upper-layer UC (leader) and middle-layer DR providers (followers) is solved using an iterative method.

\subsection{Stackelberg Game between the DR Provider and EUs}
This Stackelberg game can be formulated as the following bi-level programming problem.
\begin{align}
    \smash{\displaystyle \max_{\lambda^t_{{EU}_{ij}} \in [0,\lambda^t_{DR_i}]}} \; \sum \limits_{t \in T} \sum \limits_{j \in J} (\lambda^t_{{DR}_i} - \lambda^t_{{EU}_{ij}}) \cdot P_{{dr}_{ij}}^t    & & \forall  i \in I
\label{OF - ML - 1}
\end{align}
subject to: 
\begin{align}
(\ref{OF_EU}): (\underline{\mu_{ij}^t},\overline {\mu_{ij}^t}) & & \forall  t \in T, \forall i \in I, \forall j \in J
\end{align}
where $(\underline{\mu_{ij}^t},\overline {\mu_{ij}^t})$ is the pair of dual variables pertaining to the lower/upper bound constraints of each EU's lower-layer optimization problem in (\ref{OF_EU}).

The maximization problem in (\ref{OF_EU}) for each lower-level EU contains a concave objective function with linear inequality constraints. Therefore, it can be converted to a convex minimization problem whose global optimal solution is characterized by the Karush–Kuhn–Tucker (KKT) conditions \cite{gabriel2012complementarity}. Utilizing the KKT conditions for all the EUs and the big M method \cite{fortuny1981representation}, this bi-level optimization problem is converted to a single level problem as follows.
\begin{multline}
    \smash{\displaystyle \max_{\substack{\lambda^t_{{EU}_{ij}} \in [0,\lambda_{{DR}_i}^t], P_{{dr}_{ij}}^t \\ \overline{\mu_{ij}^t}, \underline{\mu_{ij}^t}, \xi_{ij}^t, \psi_{ij}^t}}} \;  \sum \limits_{t \in T} \sum \limits_{j \in J} (\lambda^t_{{DR}_i} - \lambda^t_{{EU}_{ij}}) \cdot P_{{dr}_{ij}}^t \\ \forall  i \in I
\label{OF-BL-1}
\end{multline}
subject to:
\begin{multline}
\lambda_{EU_{ij}}^t - \left(\frac{P_{{dr-max}_{ij}}^t}{(P_{{dr-max}_{ij}}^t-P_{{dr}_{ij}}^t)^2}\right) + \underline{\mu_{ij}^t} - \overline {\mu_{ij}^t} = 0 \\ \forall  t \in T, \forall i \in I, \forall j \in J 
\label{stationary}
\end{multline}
\begin{align}
0 \leq P_{{dr}_{ij}}^t \leq \psi_{ij}^t \cdot M_{ij}^t && \forall  t \in T, \forall i \in I, \forall j \in J 
\label{complementarity - 3}
\end{align}
\begin{align}
0 \leq \underline{\mu_{ij}^t} \leq (1-\psi_{ij}^t) \cdot M_{ij}^t && \forall  t \in T, \forall i \in I, \forall j \in J
\label{complementarity - 4}
\end{align}
\begin{multline}
0 \leq (P_{{dr-max}_{ij}}^t-P_{{dr}_{ij}}^t) \leq \xi_{ij}^t \cdot M_{ij}^t \\ \forall  t \in T, \forall i \in I, \forall j \in J
\label{complementarity - 5}
\end{multline}
\begin{align}
0 \leq \overline{\mu_{ij}^t} \leq (1-\xi_{ij}^t) \cdot M_{ij}^t && \forall  t \in T, \forall i \in I, \forall j \in J
\label{complementarity - 6}
\end{align}
\begin{align}
\psi_{ij}^t, \xi_{ij}^t \in \{0,1\} && \forall  t \in T, \forall i \in I, \forall j \in J
\label{complementarity - 7}
\end{align}
where $\psi_{ij}^t$ and $\xi_{ij}^t$ are binary variables and $M_{ij}^t$ is a large constant. The constant $M_{ij}^t$ for each EU is chosen to be the upper bound of the EU's DR quantity provision as follows.
\begin{align}
M_{ij}^t = P_{{dr-max}_{ij}}^t   && \forall  t \in T, \forall i \in I, \forall j \in J 
\label{complementarity - 8}
\end{align}

The single-level objective function (\ref{OF-BL-1}) contains a bi-linear term $\lambda_{EU_{ij}}^t \cdot P_{dr_{ij}}^t$. To eliminate this bi-linear term, the complementary slackness conditions for the constraints of the lower-level problem in (\ref{OF_EU}) are developed as follows.
\begin{align}
    P_{{dr}_{ij}}^t \cdot \underline{\mu_{ij}^t} = 0 & & \forall  t \in T, \forall i \in I, \forall j \in J
\label{complementary slackness - 1}
\end{align}
\begin{multline}
    (P_{{dr-max}_{ij}}^t-P_{{dr}_{ij}}^t) \cdot \overline{\mu_{ij}^t} = 0 
    \\  \Rightarrow   P_{{dr}_{ij}}^t \cdot \overline{\mu_{ij}^t} = P_{{dr-max}_{ij}}^t \cdot \overline{\mu_{ij}^t} \quad
    \forall  t \in T, \forall i \in I, \forall j \in J 
\label{complementary slackness - 2}
\end{multline}

By substituting (\ref{stationary}), (\ref{complementary slackness - 1})-(\ref{complementary slackness - 2}) into the bi-linear term in (\ref{OF-BL-1}), the single-level optimization problem is re-written as follows.
\begin{multline}
	\smash{\displaystyle \max_{ P_{{dr}_{ij}}^t, \overline{\mu_{ij}^t}, \underline{\mu_{ij}^t}, \xi_{ij}^t, \psi_{ij}^t}} \; \sum \limits_{t \in T} \sum \limits_{j \in J} \Bigg( \lambda^t_{{DR}_i} \cdot P_{{dr}_{ij}}^t \Bigg. \\
	\left. - \frac{P_{{dr-max}_{ij}}^t \cdot P_{{dr}_{ij}}^t}{(P_{{dr-max}_{ij}}^t-P_{{dr}_{ij}}^t)^2} - \overline {\mu_{ij}^t} \cdot P_{{dr-max}_{ij}}^t \right) \forall  i \in I
	\label{OF-BL-3}
\end{multline}
subject to:
\begin{align}
	(\ref{complementarity - 3})-(\ref{complementarity - 8}) & & \forall i \in I, \forall j \in J
	\label{eqn_24}
\end{align}

By substituting (\ref{stationary}) in the objective function (\ref{OF-BL-1}), constraint (\ref{complementarity - 4}) is no longer active because the decision variable $\underline{\mu_{ij}^t}$ no longer appears in the objective function/constraints, therefore no longer impacts the optimization problem. By removing this constraint, the only acceptable value for the binary decision variable $\psi_{ij}^t$ in (\ref{complementarity - 3}) would be 1 in order to maintain the DR quantity of the EUs within their lower bound and upper bound as presented in (\ref{OF_EU}). On the other hand, $P_{{dr}_{ij}}$ will never reach $P_{{dr-max}_{ij}}$ because the EU's inconvenience cost will surpass the revenue if $P_{{dr}_{ij}}=P_{{dr-max}_{ij}}$ (inconvenience cost will be $+\infty$, see Fig.~\ref{Inconvenience Cost}), and the DR provider will not pick up an EU whose profit is non-positive (see (\ref{OF_EU})). As a result, the binary decision variable $\xi_{ij}^t$ in (\ref{complementarity - 5}) must be 1. By replacing $\xi_{ij}^t = 1$ in (\ref{complementarity - 6}), it can be concluded that $\overline{\mu_{ij}^t} = 0$. Thus, the optimization problem can finally be simplified as follows.
\begin{multline}
    \smash{\displaystyle \max_{P_{{dr}_{ij}}^t \in [0, P_{{dr-max}_{ij}}^t]}} \; \sum \limits_{t \in T} \sum \limits_{j \in J} \Bigg( \lambda^t_{{DR}_i} \cdot P_{{dr}_{ij}}^t \Bigg. \\
	\left. - \frac{P_{{dr-max}_{ij}}^t \cdot P_{{dr}_{ij}}^t}{(P_{{dr-max}_{ij}}^t-P_{{dr}_{ij}}^t)^2} \right) \forall  i \in I
	\label{OF-BL-4}
\end{multline}

By comparing (\ref{OF-BL-4}) and (\ref{OF-BL-1}), it can be observed that $\lambda^t_{{EU}_{ij}}=\frac{P_{{dr-max}_{ij}}^t}{(P_{{dr-max}_{ij}}^t-P_{{dr}_{ij}}^t)^2}$. The above single-level maximization problem contains a concave objective function in (\ref{OF-BL-4}). The corresponding optimization problem is a convex minimization problem and can be solved in AMPL by CONOPT solver \cite{CONOPT}. In the competition among the lower-layer EUs through the non-cooperative simultaneous game, all the EUs' actions take place at the same time and players are not aware of the other players' DR provision strategies (willingness parameters) \cite{prisner2014game}. Equation (\ref{OF-BL-4}) captures the non-cooperative simultaneous game among the EUs, as the EUs compete for DR quantity provision given their $\alpha$, strategy, reported to the corresponding DR provider. 

\subsection{Stackelberg Game between the UC and The DR Providers}
This Stackelberg game is solved by the iterative approach in Algorithm 1. In each iteration, the UC sends its price signals $\lambda^t_{{DR}_i}$ (gradually increasing from 0 with the step size of 0.01 cent over the iterations) to $i^{th}$ DR provider (see Fig. \ref{Schematic Diagram}). The DR providers then solve (\ref{OF-BL-4}) to determine each EU's optimal DR quantity provision $P_{{dr}_{ij}}^t$, considering the price signal $\lambda^t_{{DR}_i}$ received from the UC. Each DR provider then sends its aggregated DR quantity $P_{{DR}_i}^t$ to the UC. This process continues until the UC as the major leader of the three-layer game finds its optimal prices to the DR providers (when the termination criteria in Algorithm \ref{Schematic Diagram} is met). These optimal prices and the associated aggregated DR quantity from the DR providers maximize the UC's net profit for utilizing DR resources. This Stackelberg game is not solved by converting the bi-level problem to a single-level problem via OPcOP, due to the non-convexity involved in the conversion process of this Stackelberg game.

\begin{algorithm} [!t]
\caption{Iterative Algorithm}\label{alg}
\begin{algorithmic} [1]
\newcommand{\Break}{\State \textbf{break} }
\State Let $iter$ and $\epsilon=0.001$ be iteration index and convergence threshold of the iterative method, respectively.
\State {The EUs report their $\alpha$ to the corresponding DR Provider.}
\State {$iter = 1$. Set $R^1_{UC}=0$} (The UC's profit at the 1st iteration is zero).
\For {$\lambda^t_{{DR}_i} \gets 0$ to $\infty$}
\State{$iter \gets iter + 1$}
\State $\text{Solve (\ref{OF-BL-4})}$ to obtain $P_{dr_{ij}}^t$ and $\lambda_{EU_{ij}}^t$
\State \Return {$P_{{DR}_i}^t=\sum \limits_{j \in J} P_{{dr}_{ij}}^t$}
\State Solve (\ref{OF Utility}) to obtain $R_{UC}^{iter}$
\If{$\mid R_{UC}^{iter} - R_{UC}^{iter-1} \mid \leq \epsilon$}
    \Break
\EndIf
\EndFor
\end{algorithmic}
\end{algorithm}

\section{Case Studies}
Case studies are performed on the IEEE 34-bus \cite{PES} and IEEE 69-bus \cite{baran1989optimal} distribution test feeders. Both residential and business DR programs are modeled. The EUs in the residential and business DR programs are selected based on their kW consumption. Time-of-use retail rates at Salt River Project (SRP) \cite{SRPResidential,SRPBusiness} are adopted as the retail rates for the residential and business EUs for their net power consumption (after DR provision). The load data is shifted such that it follows the pattern with the peak hours, off-peak hours, and super off-peak hours based on SRP's time-of-use rate program in summer. The UC's cost function coefficients $c_1$ and $c_2$ are interpolated based on three load values and their corresponding prices, following the approach in \cite{tsaousoglou2020truthful}. Two scenarios are studied for each test system to investigate the interactions among different entities and competition among the EUs via the non-cooperative simultaneous game, when the EUs adjust their parameters for DR provision willingness/strategies. 

\subsection{IEEE 34-bus Test System}

\begin{figure*}[!t]
	\centering
	\includegraphics[width=0.93\textwidth]{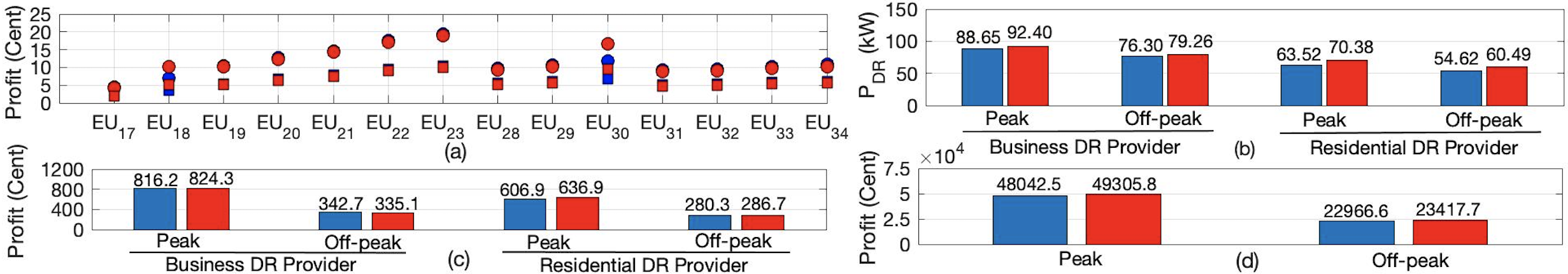}
	\caption{Profit of the EUs, aggregated DR quantity provision, profit of the DR providers, and profit of the UC in IEEE 34 bus system; The blue and red colors denote the results for scenarios 1 and 2, respectively. In (a), the data points in circle and square denote the results at peak and off-peak hours, respectively.}
	\vspace{-0.5cm}
	\label{figs-IEEE 34-bus}
\end{figure*}

\begin{table}[!t]
  \centering
  \setlength{\tabcolsep}{0.7pt}
  \caption{Optimal DR Provision and Price Signals to the EUs in IEEE 34 Bus Test System}
    \begin{tabular}{lccccccccccc}
    \toprule
    \multirow{3}[6]{*}{} & \multicolumn{5}{c}{\textbf{Scenario 1}} &       & \multicolumn{5}{c}{\textbf{Scenario 2}} \\
\cmidrule{2-6}\cmidrule{8-12}          & \multicolumn{2}{c}{\textbf{DR (kW)}} &       & \multicolumn{2}{c}{\textbf{$\lambda_{EU}$ (c/kWh)}} &       & \multicolumn{2}{c}{\textbf{DR (kW)}} &       & \multicolumn{2}{c}{\textbf{$\lambda_{EU}$(c/kWh)}} \\
\cmidrule{2-3}\cmidrule{5-6}\cmidrule{8-9}\cmidrule{11-12}          & \textbf{Off-peak} & \textbf{ Peak } &       & \textbf{Off-peak} & \textbf{ Peak } &       & \textbf{Off-peak} & \textbf{ Peak } &       & \textbf{Off-peak} & \textbf{ Peak } \\
    \midrule
    \textbf{EU 17} & 2.67  & 3.28  &       & 1.35  & 2.00  &       & 2.63  & 3.26  &       & 1.31  & 1.96 \\
    \textbf{EU 18} & 4.89  & 5.85  &       & 1.11  & 1.66  &       & 8.32  & 9.81  &       & 0.90  & 1.37 \\
    \textbf{EU 19} & 8.38  & 9.84  &       & 0.94  & 1.40  &       & 8.32  & 9.81  &       & 0.90  & 1.37 \\
    \textbf{EU 20} & 10.78 & 12.57 &       & 0.86  & 1.29  &       & 10.71 & 12.53 &       & 0.83  & 1.26 \\
    \textbf{EU 21} & 13.22 & 15.32 &       & 0.81  & 1.21  &       & 13.13 & 15.28 &       & 0.78  & 1.18 \\
    \textbf{EU 22} & 16.93 & 19.50 &       & 0.74  & 1.11  &       & 16.83 & 19.45 &       & 0.72  & 1.09 \\
    \textbf{EU 23} & 19.43 & 22.30 &       & 0.71  & 1.07  &       & 19.32 & 22.25 &       & 0.68  & 1.04 \\
    \textbf{EU 28} & 7.38  & 8.53  &       & 1.08  & 1.52  &       & 7.30  & 8.48  &       & 1.02  & 1.46 \\
    \textbf{EU 29} & 8.22  & 9.47  &       & 1.04  & 1.47  &       & 8.13  & 9.42  &       & 0.99  & 1.41 \\
    \textbf{EU 30} & 9.49  & 10.89 &       & 0.99  & 1.40  &       & 15.86 & 18.06 &       & 0.79  & 1.14 \\
    \textbf{EU 31} & 6.66  & 7.83  &       & 1.12  & 1.56  &       & 6.58  & 7.78  &       & 1.06  & 1.51 \\
    \textbf{EU 32} & 6.92  & 8.13  &       & 1.10  & 1.54  &       & 6.84  & 8.08  &       & 1.05  & 1.49 \\
    \textbf{EU 33} & 7.71  & 9.03  &       & 1.06  & 1.49  &       & 7.62  & 8.98  &       & 1.01  & 1.44 \\
    \textbf{EU 34} & 8.24  & 9.63  &       & 1.04  & 1.46  &       & 8.15  & 9.58  &       & 0.99  & 1.41 \\
    \bottomrule
    \end{tabular}%
  \label{DR Provision and Price Signal - IEEE 34-bus}%
\end{table}%

\begin{table}[!t]
  \centering
  \setlength{\tabcolsep}{3.5pt}
  \caption{Optimal Price Signals $\lambda_{DR}$ (c/kwh) from the UC to the DR providers in IEEE 34 Bus Test System}
    \begin{tabular}{llccrcc}
    \toprule
          &       & \multicolumn{2}{c}{\textbf{Scenario 1}} &       & \multicolumn{2}{c}{\textbf{Scenario 2}} \\
\cmidrule{3-4}\cmidrule{6-7}    \textbf{From} & \textbf{To} & Off-peak & Peak  &       & Off-peak & Peak \\
    \midrule
    \textbf{UC} & \textbf{Business DR provider} & 5.32  & 10.45 &       & 5.02  & 10.13 \\
    \textbf{UC} & \textbf{Residential DR provider} & 6.19  & 11.04 &       & 5.70  & 10.42 \\
    \bottomrule
    \end{tabular}%
  \label{tab:Price Signals from UC to the DR providers - IEEE 34}%
\end{table}%

In this test system, the residential DR program includes the residential EUs 28-34. The residential EUs 28-30 have identical load profile and their base load is 75 kW. The residential EUs 31-34 have identical load profile and their base load is 57 kW. The business DR program includes the business EUs 17-23. All these business EUs have identical load profile and their base load is 230 kW. The UC's cost function coefficients $c_1$ and $c_2$ are -1088.2 and 0.2024, respectively. In scenario 1, 1) the willingness parameters $\alpha$ for business EUs 17-23 are 0.03, 0.05, 0.08, 0.10, 0.12, 0.15, and 0.17, respectively; 2) the willingness parameters for residential EUs 28-34 are 0.20, 0.22, 0.25, 0.29, 0.30, 0.33, and 0.35, respectively. In scenario 2, 1) the business EU 18 and residential EU 30 increase their willingness parameters from 0.05 to 0.08 and from 0.25 to 0.40, respectively; 2) the willingness parameters of all the other EUs remain unchanged from scenario 1. Table \ref{DR Provision and Price Signal - IEEE 34-bus} depicts 1) the optimal DR provision of each business/residential EU; and 2) the optimal prices $\lambda_{EU}$ from each business/residential DR provider to each EU. Table \ref{tab:Price Signals from UC to the DR providers - IEEE 34} depicts the optimal prices $\lambda_{DR}$ from the UC to each DR provider. Figs. (\ref{figs-IEEE 34-bus}a)-(\ref{figs-IEEE 34-bus}d) show the profit of each EU, the aggregated DR quantity purchased by the DR providers, the profit of the DR providers, and the profit of the UC, during peak hours and off-peak hours, respectively. Table \ref{DR Provision and Price Signal - IEEE 34-bus} and Fig. (\ref{figs-IEEE 34-bus}a) indicate that among the EUs with identical load profile in the same DR program, the EU with greater DR provision willingness (greater $\alpha$) contributes more in the DR program and earns greater profit. The competition among the EUs in the same DR program (via the non-cooperative simultaneous game) during peak and off-peak hours can be observed in Table \ref{DR Provision and Price Signal - IEEE 34-bus} and Fig. (\ref{figs-IEEE 34-bus}a) by comparing the results from scenarios 1 and 2. Among the business EUs, the EU 18, whose DR provision willingness increased from scenario 1 to scenario 2, contributed more in the DR program and earned more profit in scenario 2. As the EU 18 changes its DR strategy/willingness, the DR contribution and profit of other EUs in the business DR program decreased from scenario 1 to scenario 2. Similarly, among the residential EUs, the EU 30, whose DR provision willingness increased from scenario 1 to scenario 2, contributed more in the DR program and earned more profit in scenario 2. As the EU 30 changes its DR strategy/willingness, the DR contribution and profit of other EUs in the residential DR program decreased from scenario 1 to scenario 2. Fig. (\ref{figs-IEEE 34-bus}b) indicates that as the EU 18 is willing to sell more DR quantity, the aggregated DR quantity purchased by the business DR provider increases from scenario 1 to scenario 2 during both peak and off-peak hours, and the prices from/to the business DR provider consequently decrease from scenario 1 to scenario 2 (in Tables \ref{DR Provision and Price Signal - IEEE 34-bus} and \ref{tab:Price Signals from UC to the DR providers - IEEE 34}). For the residential DR provider, when the EU 30 is willing to sell more DR quantity, similar trends can be observed with the increased purchase of the aggregated DR quantity and decreased price signals from the UC and to the residential EUs during peak and off-peak hours. Figs. (\ref{figs-IEEE 34-bus}c)-(\ref{figs-IEEE 34-bus}d) show the residential DR provider and the UC earned greater profit in scenario 2 compared to scenario 1, since the greater willingness for DR quantity provision in scenario 2 enabled them to purchase more DR quantity at lower prices. However, while the business DR provider's profit increased during peak hours from scenario 1 to scenario 2, this DR provider's profit decreased during off-peak hours from scenario 1 to scenario 2 in spite of purchasing greater DR quantity during off-peak hours in scenario 2. This is because even though the UC aims to run the DR program to reduce its operation cost, the UC is not interested in purchasing more DR quantity than needed. Purchasing DR quantity causes electricity bills revenue reduction for the UC (see (\ref{Revenue - EB})). If the DR providers purchase more DR quantity from the EUs than the UC needs, the incentive they receive drops in a way that their profit decreases overall (see (\ref{OF - ML})).

\subsection{IEEE 69-bus Test System}

\begin{figure*}[!t]
	\centering
	\includegraphics[width=0.93\textwidth]{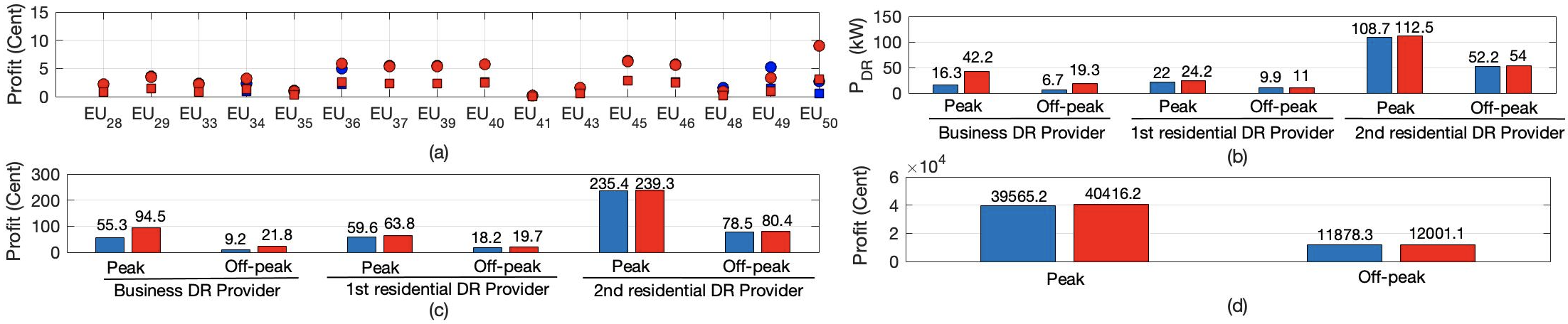}
	\caption{Profit of the EUs, aggregated DR quantity provision, profit of the DR providers, and profit of the UC in IEEE 69 bus system; The blue and red colors denote the results for scenarios 1 and 2, respectively. In (a), the data points in circle and square denote the results at peak and off-peak hours, respectively.}
	\vspace{-0.5cm}
	\label{figs-IEEE 69-bus}
\end{figure*}

\begin{table}[!t]
  \centering
  \setlength{\tabcolsep}{0.95pt}
  \caption{Optimal DR Provision and Price Signals to the EUs in IEEE 69 Bus Test System}
    \begin{tabular}{lccccccccrcc}
    \toprule
          & \multicolumn{5}{c}{\textbf{Scenario 1}} &       & \multicolumn{5}{c}{\textbf{Scenario 2}} \\
\cmidrule{2-6}\cmidrule{8-12}          & \multicolumn{2}{c}{\textbf{DR (kW)}} &       & \multicolumn{2}{c}{\textbf{$\lambda_{EU}$ (c/kWh)}} &       & \multicolumn{2}{c}{\textbf{DR (KW)}} &       & \multicolumn{2}{c}{\textbf{$\lambda_{EU}$ (c/kWh)}} \\
\cmidrule{2-3}\cmidrule{5-6}\cmidrule{8-9}\cmidrule{11-12}          & \textbf{Off-peak} & \textbf{Peak} &       & \textbf{off-peak} & \textbf{Peak} &       & \textbf{Off-peak} & \textbf{ Peak } &       & \textbf{ off-peak } & \textbf{ Peak } \\
    \midrule
    \multicolumn{1}{c}{\textbf{EU 28}} & 1.88  & 4.21  &       & 0.959 & 0.892 &       & 1.86  & 4.18  &       & 0.940 & 0.873 \\
    \multicolumn{1}{c}{\textbf{EU 29}} & 3.44  & 7.35  &       & 0.796 & 0.746 &       & 3.41  & 7.31  &       & 0.780 & 0.730 \\
    \multicolumn{1}{c}{\textbf{EU 33}} & 1.90  & 4.24  &       & 0.957 & 0.890 &       & 1.88  & 4.21  &       & 0.938 & 0.871 \\
    \multicolumn{1}{c}{\textbf{EU 34}} & 2.01  & 4.47  &       & 0.940 & 0.875 &       & 3.15  & 6.78  &       & 0.800 & 0.748 \\
    \multicolumn{1}{c}{\textbf{EU 35}} & 0.70  & 1.74  &       & 1.285 & 1.177 &       & 0.69  & 1.72  &       & 1.260 & 1.153 \\
    \multicolumn{1}{c}{\textbf{EU 36}} & 7.11  & 14.86 &       & 0.509 & 0.484 &       & 9.17  & 18.90 &       & 0.464 & 0.442 \\
    \multicolumn{1}{c}{\textbf{EU 37}} & 8.05  & 16.71 &       & 0.489 & 0.466 &       & 8.03  & 16.67 &       & 0.484 & 0.460 \\
    \multicolumn{1}{c}{\textbf{EU 39}} & 8.01  & 16.62 &       & 0.490 & 0.467 &       & 7.98  & 16.58 &       & 0.485 & 0.461 \\
    \multicolumn{1}{c}{\textbf{EU 40}} & 8.71  & 18.00 &       & 0.477 & 0.455 &       & 8.68  & 17.95 &       & 0.472 & 0.449 \\
    \multicolumn{1}{c}{\textbf{EU 41}} & 0.11  & 0.47  &       & 1.555 & 1.390 &       & 0.10  & 0.46  &       & 1.543 & 1.375 \\
    \multicolumn{1}{c}{\textbf{EU 43}} & 1.65  & 3.87  &       & 0.799 & 0.746 &       & 1.64  & 3.85  &       & 0.792 & 0.737 \\
    \multicolumn{1}{c}{\textbf{EU 45}} & 9.84  & 20.21 &       & 0.458 & 0.438 &       & 9.81  & 20.16 &       & 0.454 & 0.432 \\
    \multicolumn{1}{c}{\textbf{EU 46}} & 8.68  & 17.94 &       & 0.477 & 0.455 &       & 8.65  & 17.90 &       & 0.473 & 0.450 \\
    \multicolumn{1}{c}{\textbf{EU 48}} & 0.83  & 2.39  &       & 1.003 & 1.210 &       & 0.69  & 2.10  &       & 0.837 & 0.913 \\
    \textbf{EU 49} & 4.17  & 9.62  &       & 0.620 & 0.774 &       & 3.82  & 8.95  &       & 0.512 & 0.578 \\
    \textbf{EU 50} & 1.68  & 4.30  &       & 0.819 & 1.004 &       & 14.77 & 31.15 &       & 0.334 & 0.385 \\
    \bottomrule
    \end{tabular}%
  \label{DR Provision and Price Signal - IEEE 69-bus}%
\end{table}%

\begin{table}[!t]
  \centering
  \setlength{\tabcolsep}{2pt}
  \caption{Optimal Price Signals $\lambda_{DR}$ (c/kwh) from the UC to the DR providers in IEEE 69 Bus Test System}    \begin{tabular}{llccccc}
    \toprule
          &       & \multicolumn{2}{c}{\textbf{Scenario 1}} &       & \multicolumn{2}{c}{\textbf{Scenario 2}} \\
\cmidrule{3-4}\cmidrule{6-7}    \textbf{From} & \textbf{To} & \textbf{Off-peak} & \textbf{Peak} &       & \textbf{Off-peak} & \textbf{Peak} \\
    \midrule
    \textbf{UC} & \textbf{Business DR provider} & 2.09  & 4.29  &       & 1.52  & 2.69 \\
    \textbf{UC} & \textbf{1st Residential DR provider} & 2.75  & 3.57  &       & 2.66  & 3.45 \\
    \textbf{UC} & \textbf{2nd Residential DR provider} & 2.00  & 2.64  &       & 1.97  & 2.59 \\
    \bottomrule
    \end{tabular}%
  \label{tab:Price Signals from UC to the DR providers - IEEE 69}%
\end{table}%

In this test system, the 1st and 2nd residential DR programs include the residential EUs 28-35 and EUs 36-46, respectively. The business DR program includes the business EUs 47-50. There is no load at buses/EUs 30, 31, 32, buses/EUs 38, 42, 44, and bus/EU 47 in the territories of the 1st and 2nd residential DR providers and the business DR provider, respectively. In the residential DR program 1, the base load for EUs 28-35 is 26 kW, 26 kW, 14 kW, 19.5 kW, and 6 kW, respectively. In the residential DR program 2, the base load for EUs 36-46 is 26 kW, 26 kW, 24 kW, 24 kW, 1.2 kW, 6 kW, 39.22 kW, and 39.22 kW, respectively. In the business DR program, the base load for EUs 48-50 is 79 kW, 384.7 kW, and 384.7 kW, respectively. The UC's cost function coefficients $c_1$ and $c_2$ are -14.3 and 0.004506, respectively. In scenario 1, 1) the willingness parameters $\alpha$, in the 1st residential DR program, for EUs 28-35 are 0.15, 0.24, 0.28, 0.21, and 0.32, respectively; 2) the willingness parameters, in the 2nd residential DR program, for EUs 36-46 are 0.46, 0.51, 0.55, 0.59, 0.7, 0.64, 0.4, and 0.36, respectively; 3) the willingness parameters, in the business DR program, for EUs 48-50 are 0.03, 0.02, and 0.01, respectively. In scenario 2, 1) the EU 34 (in the 1st residential DR program), EU 36 (in the 2nd residential DR program), and EU 50 (in the business DR program) increase their willingness parameters from 0.21 to 0.30, from 0.46 to 0.57, and from 0.01 to 0.06, respectively; 2) the willingness parameters of all the other EUs remain unchanged from scenario 1. Table \ref{DR Provision and Price Signal - IEEE 69-bus} depicts 1) the optimal DR provision of each business/residential EU; and 2) the optimal prices $\lambda_{EU}$ from each business/residential DR provider to each EU. Table \ref{tab:Price Signals from UC to the DR providers - IEEE 69} depicts the optimal prices $\lambda_{DR}$ from the UC to each DR provider. Figs. (\ref{figs-IEEE 69-bus}a)-(\ref{figs-IEEE 69-bus}d) show the profit of each EU, the aggregated DR quantity purchased by the DR providers, the profit of the DR providers, and the profit of the UC, during peak hours and off-peak hours, respectively. Table \ref{DR Provision and Price Signal - IEEE 69-bus} and Fig. (\ref{figs-IEEE 69-bus}a) indicate that among the EUs with identical load profile in the same DR program, the EU with greater DR provision willingness (greater $\alpha$) contributes more in the DR program and earns greater profit. The competition among the EUs in the same DR program (via the non-cooperative simultaneous game) during peak and off-peak hours can be observed in Table \ref{DR Provision and Price Signal - IEEE 69-bus} and Fig. (\ref{figs-IEEE 69-bus}a) by comparing the results from scenarios 1 and 2. Among the EUs in the 1st residential DR program, the EU 34, whose DR provision willingness increased from scenario 1 to scenario 2, contributed more in the DR program and earned more profit in scenario 2. As the EU 34 changes its DR strategy/willingness, the DR contribution and profit of other EUs in the 1st residential DR program decreased from scenario 1 to scenario 2. Similar trend can be observed in the 2nd residential DR program and the business DR program where the EU 36 and the EU 50  increased their DR provision willingness parameters from scenario 1 to scenario 2, respectively. Fig. (\ref{figs-IEEE 69-bus}b) indicates that as the EU 34 is willing to sell more DR quantity, the aggregated DR quantity purchased by the 1st residential DR provider increases from scenario 1 to scenario 2 during both peak and off-peak hours, and the prices from/to the 1st residential DR provider consequently decrease from scenario 1 to scenario 2 (in Tables \ref{DR Provision and Price Signal - IEEE 69-bus} and \ref{tab:Price Signals from UC to the DR providers - IEEE 69}). For the 2nd residential DR provider (or the business DR provider), when the EU 36 (or EU 50) is willing to sell more DR quantity, similar trends can be observed with the increased purchase of the aggregated DR quantity and decreased price signals from the UC to the 2nd residential EUs (or the business EUs) during peak and off-peak hours. Figs. (\ref{figs-IEEE 69-bus}c)-(\ref{figs-IEEE 69-bus}d) show the DR providers and the UC earned greater profit in scenario 2 compared to scenario 1, since the greater willingness for DR quantity provision in scenario 2 enabled them to purchase more DR quantity at lower prices.

\section{Conclusion}
This paper proposed a framework with three-layer coupled games to study the optimal dispatch and incentivization of third-party DR resources in the vertically integrated UC. The interactions and competitions among the profit-seeking process of the UC, the third-party DR providers, and the individual EUs in the DR programs are investigated. It was shown that the UC, the DR providers, and the EUs can increase their net benefit/profit through the third-party DR programs by imposing appropriate pricing and DR provision decisions. Through this framework, the overall net profit for running the DR programs was optimally allocated among the UC, the third-party DR providers, and the EUs, by enabling these entities to jointly determine the optimal price signals and DR provision quantity considering the benefit of all these involved entities. In the case studies, the competitions among the EUs in each DR program via the simultaneous game were demonstrated. When certain EU changed its DR provision strategy/willingness by adjusting the willingness coefficient in the proposed EU inconvenience cost model, this EU could increase its net benefit/profit while reducing the net benefit/profit of other EUs in the same DR program. Future work could be 1) adopting more computationally efficient solution methods for solving the game between the UC and the DR providers such that the method can be utilized for both offline studies and real-time operations; and 2) enhancing the coupled game model by allowing the EUs to adjust their DR willingness parameters strategically based on the price signals.

\bibliographystyle{IEEEtran}
\bibliography{IEEEfull}

\end{document}